\documentclass[aps, prl, reprint, groupedaddress, superscriptaddress]{revtex4-1}
\usepackage{amsmath,amssymb}
\usepackage{graphicx}
\usepackage{color}
\usepackage{bm}
\usepackage[colorlinks,linkcolor=blue]{hyperref}

\begin{document}
\title{Microscopic Marangoni flows cannot be predicted on the basis of pressure gradients}\label{title}

\author{Yawei Liu}
\affiliation{Beijing Advanced Innovation Center for Soft Matter Science and Engineering, Beijing University of Chemical Technology, Beijing 100029, China}
\affiliation{State Key Laboratory of Organic-Inorganic Composites, Beijing University of Chemical Technology, Beijing 100029, China}
\author{Raman Ganti}
\affiliation{Department of Chemistry, University of Cambridge, Lensfield Road, Cambridge CB2 1EW, UK}
\author{Hugh G. A. Burton}
\affiliation{Department of Chemistry, University of Cambridge, Lensfield Road, Cambridge CB2 1EW, UK}
\author{Xianren Zhang}
\affiliation{State Key Laboratory of Organic-Inorganic Composites, Beijing University of Chemical Technology, Beijing 100029, China}
\author{Wenchuan Wang}
\affiliation{Beijing Advanced Innovation Center for Soft Matter Science and Engineering, Beijing University of Chemical Technology, Beijing 100029, China}
\author{Daan Frenkel}
\email[Corresponding author: ]{df246@cam.ac.uk}
\affiliation{Department of Chemistry, University of Cambridge, Lensfield Road, Cambridge CB2 1EW, UK}

\date{\today}

\begin{abstract}
	
A concentration gradient along a fluid-fluid interface can cause flow. On a microscopic level, this so-called  Marangoni effect can be viewed as being caused by  a gradient in the pressures acting on the fluid elements, or as the chemical-potential gradients acting on the excess densities of different species at the interface. If the interfacial thickness can be ignored, all approaches should result in the same flow profile away from the interface. However, on a more microscopic scale, the different expressions result in different flow profiles, only one of which can be correct. Here we compare the results of direct non-equilibrium Molecular Dynamics simulations with the flows that would be generated by pressure and  chemical potential gradients. We find that the approach based on the chemical potential gradients agrees with the direct simulations, whereas the calculations based on the pressure gradients do not. 

\end{abstract}

\maketitle
Fluid flows can be generated by variations of temperature or solute concentration parallel to a fluid-fluid interface. This phenomenon is  known as the Marangoni effect (see e.g.~\cite{Loewenthal1931,Hershey1939}). The `continuum'  explanation for this effect is that the gradients in temperature or concentration result in gradients in the surface tension, which then induce shear flow~\cite{Young1959,Sternling1959,Levich1962,Levich1969,Ruckenstein1981,Anderson1989,Fanton1998}. 
For some applications, it would  be interesting to have a higher-resolution description of Marangoni flows. The reason is that the local shear in Marangoni flows can be quite large and could become important for nano-fluidics~\cite{Stone2004,Squires2005,Bocquet2010a,Bocquet2014}. Moreover, the precise Marangoni flow profile might affect the orientation and even conformation of molecules such as proteins near an interface. At present, such microscopic insights in Marangoni flows are lacking. 

In this paper, we  study the flow induced in  a flat liquid-liquid interface by the  concentration gradient of  a neutral solute (i.e. the `solutal' Marangoni effect). We first review the various (pressure or chemical potential-based)  expressions for the force acting on molecules near the interface. We then perform Molecular Dynamics (MD) simulations to compare the flows generated by these forces with the results of direct non-equilibrium simulations.

Let us first consider two immiscible liquids at  temperature $T$, that meet at  a flat liquid-liquid interface at $z(x,y)=0$. When a concentration gradient is applied along $x$, flow occurs due to a surface-tension gradient. The surface tension, $\gamma$, can be related to  the integral of the difference between the longitudinal and transverse pressures  near the interface:
\begin{equation}\label{eq:eq01}
\gamma=\int_{-\infty}^{+\infty} [ p^{zz}(z)-p^{xx}(z) ] dz,
\end{equation}
where  $p^{zz}(z)$ and $p^{xx}(z)$ are the normal and transverse components of the pressure tensor at $z$, respectively~\cite{Tolman1948,Kirkwood1949,Marchand2011}. 
As was pointed out by Schofield and Henderson~\cite{Schofield1982}, the integral in Eq.~\ref{eq:eq01} does not depend on the choice of the expression for the pressure tensor. Nevertheless,  the microscopic flow near an interface is expected to depend on the local gradients of the pressure, rather than the gradients  of the integral of the pressure tensor. This is  important because, close to the interface, the viscosity of the liquid need not be constant, hence making a difference where the forces act. 

The most intuitive method to obtain the Marangoni force acting on fluid molecules near the interface is to calculate the force per unit volume on a small volume element from the pressure gradient  ($\partial p^{xx}(z) / \partial x$), and then obtain the force per particle by dividing the force per volume by the local number density. We now make the assumption that $p^{xx}(z)$ depends on $x$ only through its dependence on the spatial variation in the bulk concentration $\rho_C$ (or, equivalently, the chemical potential)  of the species subject to a concentration gradient:
\begin{equation}\label{eq:eq02}
f^{V}(z) = -\frac{\partial p^{xx}(z)}{\partial \rho_C}  \frac{\partial \rho_C}{\partial x}.
\end{equation}
We note that the condition for mechanical equilibrium in the bulk implies, via the Gibbs-Duhem relation, that a concentration gradient in a `solute' also causes a gradient in the concentration of the solvent. However, these other gradients are not independent, and hence we will treat the solute concentration gradient as the independent variable.

The general expression for the local pressure tensor at  position ${\bf r}$ is given by 
\begin{multline}\label{eq:eq03}
p^{ab}({\bf r})=\rho({\bf r}) k_BT\delta^{ab} \\
+\frac{1}{V} \bigg\langle \sum_i \sum_{j>i} r_{ij}^a f_{ij}^b \xi^{ab}\left({\bf r},{\bf r}_i,{\bf r}_j\right)\bigg\rangle,
\end{multline}
under the condition of $\int d{\bf r} \xi^{ab}\left({\bf r},{\bf r}_i,{\bf r}_j\right) =1$~\cite{Irving1950,Schofield1982,Gloor2005}. Here, $a$ and $b$ denote the cartesian components of the pressure tensor, $\rho({\bf r})$ is the local density, $\delta^{ab}$ denotes the Kronecker delta, $r_{ij}$ and $f_{ij}$ represent the distance and force between particles $i$ and $j$, and $\xi^{ab}\left({\bf r},{\bf r}_i,{\bf r}_j\right) $ is the fraction of the intermolecular virial from a given pair of molecules at ${\bf r}_i$ and ${\bf r}_j$ to be assigned to position ${\bf r}$. As was argued in ref.~\onlinecite{Schofield1982}, there is no unambiguous way to assign the intermolecular virial in the system. All definitions of the pressure tensor that differ only by a function that is divergence-free  are acceptable. There are, in fact, several  widely used
definitions for the local pressure tensor~\cite{Kirkwood1949,Harasima1953}. For example, for a given pair of molecules, the virial definition specifies that half of the contribution to the stress resides in each elemental volume containing the molecule~\cite{Nijmeijer1988}, while the Irving-Kirkwood definition specifies that the contribution is evenly distributed along a line connecting the two molecules~\cite{Irving1950}. These two definitions lead to the same value of the surface tension, but to very different results for the pressure tensor distribution in the interface~\cite{Walton1983,Weng2000}. 

Gibbs was the first to give a consistent thermodynamic description of the surface tension~\cite{Gibbs1957}. In particular, Gibbs related the variation of the surface tension with chemical potential of species $i$ to the excess of that species at the interface. For an $n$-component system:  $d\gamma=-\sum_{i=1}^n \Gamma_i d \mu_i$, with $\Gamma_i$ the surface excess and $d\mu_i$ the chemical potential variation due to the concentration gradient. We assume that  ${\partial \mu_i} / {\partial x}$ is independent of $z$ (fast  equilibration normal to the interface). 
Because $\Gamma_i\equiv\int_{-\infty}^{\infty} \left( \rho_i(z,x)-\rho_i^\text{bulk}(x) \right) dz$, the surface tension gradient along $x$ is
\begin{multline}\label{eq:eq04}
\frac{\partial \gamma}{\partial x} = \int_{-\infty}^{\infty} \sum_{i=1}^n \left( \rho_i(z,x)-\rho_i^\text{bulk}(x) \right) \left( -\frac {\partial \mu_i} {\partial x} \right) dz.
\end{multline}
This suggests that the local force acting on a volume element at ${\bf r}$ is given by 
$-\sum_{i=1}^n \Gamma_i({\bf r}) \left( -{\partial \mu_i} / {\partial x} \right)$.
Such a relation also follows from the Gibbs-Duhem equation $Vdp=\sum_{i=1}^n N_i d \mu_i$ with $N_i$ the number of particles of component $i$ in volume $V$ and $p$ the pressure. Let us denote  the number density of component $i$ in the mixture by $\rho_i$. Then $dp=\sum_{i=1}^n \rho_i d \mu_i$. A concentration gradient of  component $i$ along $x$ will lead to a chemical potential gradient $\partial \mu_i / \partial x$. As the pressure remains constant in the bulk, we must have $0= \sum_{i=1}^n \rho_i^\text{bulk}(x) \left( \partial \mu_i / \partial x \right)$. At a position $z$ near the interface, a pressure gradient remains giving a force per unit volume
\begin{multline}\label{eq:eq05}
f^V(z)=\left( -\frac {\partial p(z,x)} {\partial x} \right) \\
= \sum_{i=1}^n \left( \rho_i(z,x)-\rho_i^\text{bulk}(x) \right) \left( -\frac {\partial \mu_i} {\partial x} \right).
\end{multline}
We can interpret $\left( -{\partial \mu_i} / {\partial x} \right)$ as the force  per-atom acting on the particles of component $i$. This expression is convenient, because the imposed chemical potential gradients are constant throughout the system. In the bulk, the composition is such that the forces balance (because the bulk pressure equilibrates rapidly). Upon approaching the interface, the concentration of different components may change, leading to non-zero net forces. In other words:  particles of a given species experience the same force regardless of their distance from the interface. The force acting on species $i$ is then
\begin{equation}\label{eq:eq06}
f_i = \left( -\frac {\partial \mu_i^\text{bulk}} {\partial x} \right) =\left( -\frac {\partial \mu_i^\text{bulk}} {\partial \rho_i} \right)_{P} \cdot \nabla \rho_i.
\end{equation}
We now have  two alternative expressions (Eq.~\ref{eq:eq02} and Eq.~\ref{eq:eq05}) for the surface force arising in the solutal Marangoni effect. Both satisfy that the integrated surface force is  equal to the surface tension gradient, but otherwise they are not obviously identical. 

\begin{figure}[tb]
	\begin{center}
		 \includegraphics[width=0.4\textwidth]{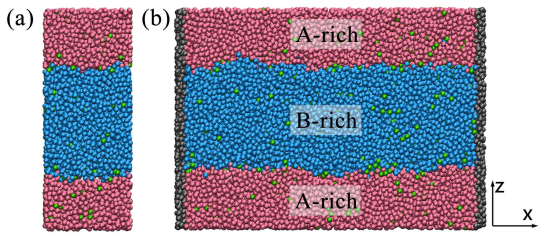}
		\caption{(a) Simulation box used to compute the force and flow profiles using the forces obtained from Eq.~\ref{eq:eq06} and Eq.~\ref{eq:fzV}. 
 (b) Simulation box used in the non-equilibrium MD simulations with explicitly imposed concentration gradients. 
 The red and blue particles represent the two solvents ($A$ and $B$), the green particles represent the solute ($C$), and the black particles represent the solid walls. }
		\label{fig:fig01}
	\end{center}
\end{figure}

To test which, if any, of these microscopic expressions is correct, we performed  MD simulations on a simple model system. We consider a fluid mixture composed of one solute ($C$) and two immiscible solvents($A$ and $B$, respectively), with two liquid-liquid interfaces, as shown in Fig.~\ref{fig:fig01}. All particles are assumed to have the same mass $m$ and molecular radius $\sigma$. They interact through Lennard-Jones potentials, $U_{\alpha\beta}(r) = 4 \epsilon_{\alpha\beta} \left[ (\sigma / r)^{12} - (\sigma / r)^6 \right]$ ($\alpha, \beta \in \left\{ A,B,C \right\}$) with interaction energy $\epsilon_{\alpha\beta}$. All interactions are truncated and shifted at $4\sigma$. For simplicity, we focus on ideal solutions composed of identical solvent and solute particles and take $\epsilon_{AA} = \epsilon_{BB} = \epsilon_{CC} = \epsilon_{AC} = \epsilon_{BC} \equiv  1.0  \epsilon$ (which defines our unit of energy). However,  $A$ and $B$ tend to demix because they have a weaker attraction:   $\epsilon_{AB}$=0.3 $\epsilon$. Throughout this article we use reduced units, with $\sigma$, $\epsilon$ and $m$ denoting the units of length, energy and mass respectively. 

All simulations were carried out using LAMMPS~\cite{Plimpton1995} in an isothermal, isobaric ($Np^{zz}T$) ensemble. Periodic boundary conditions were imposed in all directions. The temperature and normal pressure during the simulations were maintained at $T=0.846$ and $p_{ex}=0.012$. The relaxation parameter for the Nos\'e-Hoover thermostat is set to $0.1$, and that for the pressure barostat is $1$. The velocity-Verlet algorithm with a time step of $0.001$ is used for the integration of equations of motion. All simulations were run for $2\times10^8 - 4\times10^8$ steps to obtain good statistics.

The computation of $\partial p^{xx}(z)/\partial \rho_C$ requires several equilibrium simulations at  a constant bulk concentration. These can be carried out in a relatively small simulation box, shown in Figure~\ref{fig:fig01}(a). The box dimensions were  $L_x$=16.44 and $L_y$=9.86, $\langle L_z \rangle$=42.4 ($L_z$ fluctuates, and depends very weakly on the solute concentration). The system contained $5040$ particles, approximately equally distributed between the $A$-phase and the $B$-phase.
 To compute the composition-dependence of $p^{xx}(z)$, we performed simulations where we varied the concentration of the solute $C$, while keeping the total number of particles fixed.

From the numerical estimate of $\partial p^{xx}(z)/\partial \rho_C$, we computed the corresponding force at $\rho_C = 0.02$ and $\Delta \rho_C=0.01$ using
\begin{multline}\label{eq:fzV}
f^{V}(z) = -\frac{\partial p^{xx}(z)}{\partial \rho_C}  \frac{\partial \rho_C}{\partial x} \\
\approx  -\frac {p^{xx}_{\rho_C+\Delta \rho_C}(z)-p^{xx}_{\rho_C-\Delta \rho_C}(z)} {2\Delta \rho_C} \cdot \nabla \rho_C
\end{multline}
We verified that our estimate for the pressure gradient did not depend on our choice of $\Delta \rho_C$.
Subsequently, we converted the force per unit volume to a force per particle, by dividing by the total number density at height $z$, $\rho(z)$. Such a body-force might, for instance, be due to a gravitational field. These per-particle forces were then applied in a non-equilibrium simulation of the fluid with solute density  $\rho_C$ to measure the corresponding flow profile at $\nabla \rho_C$.

Starting from Eq.~\ref{eq:eq06} we can compute the forces that would result from the gradient of  the chemical potentials. These  per-atom forces were applied to the solute and the solvent particles. During all these simulations with explicit forces, a constant force is applied to all fluid particles  to balance the surface force, to ensure that there is no center-of-mass flow. To measure the local velocity, the simulation box was divided into a series of slabs of thickness $dz = 0.05$ parallel to the interface. The local velocity is computed as the time-averaged center-of-mass velocity of all fluid particles in each slab.

To compare, we performed non-equilibrium simulations where a concentration gradient was explicitly imposed.  Figure~\ref{fig:fig01}(b) shows the simulation box in which the fluid mixture has a constant bulk concentration gradient along $x$. The box size $L_x$=59.19 and $L_y$=9.86, and $\langle L_z \rangle$= 42.4. The system contained $18144$ particles, approximately equally distributed between the $A$-phase and the $B$-phase. Non-equilibrium simulations were carried out to measure the flow profile at a given value of $\nabla \rho_C$. In this case, we employed a box that  was terminated on both ends by hard walls perpendicular to the  $x$-direction. These walls were composed of frozen fluid particles that interact with the fluid via Lennard-Jones potentials where $\epsilon_{WA} = \epsilon_{WB} = \epsilon_{WC} = 1.0\epsilon$.  Next to each wall, we defined a `source' region with a width of $8$. During the simulations, every $500$ steps, the types of the fluid particles in the bulk of these source regions are reset to maintain constant bulk concentrations on the two sides and a steady gradient along $x$. In the simulation, flow in the interface, set in motion by the surface force, is accompanied by a bulk back-flow caused by the presence of the walls.
\begin{figure}[tb]
	\begin{center}
		 \includegraphics[width=0.25\textwidth]{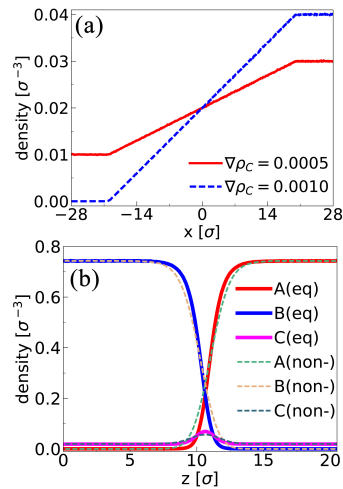}
		\caption{(a) The bulk concentration profiles along $x$ from the non-equilibrium simulations.  (b) The density profiles along $z$ near one interface at $z=10.6$ from the non-equilibrium simulation at $\nabla \rho_C =0.0005$ (dashed lines), and from the equilibrium simulation at $\rho_C\sim0.02$ (solid lines). The density profiles near another interface at $-10.6$ are the same.}
		\label{fig:fig02}
	\end{center}
\end{figure}
\begin{figure*}[tb]
	\begin{center}
		\includegraphics[width=0.75\textwidth]{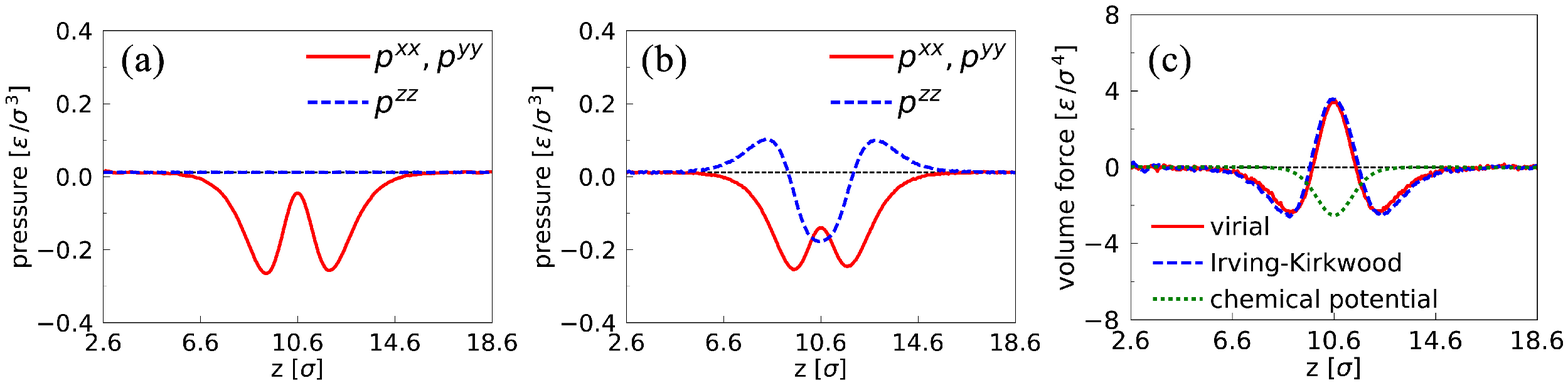}
		\caption{(a-b) The pressure profiles along $z$ near one interface at $z=10.6$ calculated by the Irving-Kirkwood definition (a) and virial definition (b) at $\rho_C\sim0.01$. The black dashed line shows the external pressure of $p_{ex}=0.012$. 
			 (c) The volume force profiles per unit concentration gradient  along $z$ near one interface at $z=10.6$ calculated by different methods at $\rho_C\sim0.02$. The horizontal dashed line corresponds to the force of zero.}
		\label{fig:fig03}
	\end{center}
\end{figure*}

Figure~\ref{fig:fig02}(a) shows the bulk concentration profiles along $x$ for  the non-equilibrium simulations. 
The figure shows that the concentration profile is linear between the limiting values imposed at the walls. 
The concentration gradients in two independent simulations were $\nabla \rho_C=0.001$ and $\nabla \rho_C=0.0005$, respectively. Figure~\ref{fig:fig02}(b) shows the density profiles for each component along $z$ from the simulation of $\nabla \rho_C=0.0005$. As the average bulk concentration is $0.02$, the density profiles from the equilibrium simulation at $\rho_C \sim 0.02$ are also plotted here for comparison. The results show good agreement for the local densities from the two simulations, indicating that in the non-equilibrium simulation, fluid states are still close to equilibrium and that equilibration along $z$ is  fast. Both are assumptions we adopted to calculate the surface force via Eq.~\ref{eq:eq02} and Eq.~\ref{eq:eq05}.

In order to calculate the surface force at $\rho_C \sim 0.02$ via Eq.~\ref{eq:eq02}, we computed the pressure-tensor profile at $\rho_C \sim 0.01$ and  $\rho_C \sim 0.03$. Figure~\ref{fig:fig03} shows the pressure profiles along $z$ near a liquid-liquid interface at $\rho_C\sim0.01$ using the Irving-Kirkwood and virial definitions. In the bulk where the fluid is homogeneous, both definitions lead to the same value since $p^{zz} = p^{xx} = p_{ex} = 0.012$. Upon approaching the interface, $p^{zz}$ from the Irving-Kirkwood definition is (necessarily) the same as the bulk pressure, reflecting mechanical equilibrium along $z$ [Fig.~\ref{fig:fig03} (a)]. As is well known the virial expression for $p^{zz}$ is not constant  [Fig.~\ref{fig:fig03} (b)]. We verified that the two expressions for the pressure tensor did yield the same value of surface tension.  We find (from Eq.~\ref{eq:eq01}) that  the surface tension is $1.14$  at $\rho_C \sim 0.01$ and $1.05$ at $\rho_C \sim 0.03$.

The chemical potential for component $i$ is given by $\mu_i = \mu_i^0 + k_BT \ln \rho_i^\text{bulk} + \mu_i^{exc}$, with $k_B$ the Boltzmann constant. $\mu_i^0$ denotes a (constant) reference value and  $\mu_i^{exc}$ denotes the excess chemical potential due to intermolecular interactions. Because the bulk solutions  are ideal, $\mu_i^{exc}$ does not depend on the concentration of $C$. Thus, at $\rho_C \sim 0.02$, with $\rho_A^\text{bulk}=\rho_B^\text{bulk}=0.742$ and $\rho_C^\text{bulk}=0.019$ [Fig.~\ref{fig:fig02}(b)], if $\nabla \rho_C=1.0$, we obtain $f_A=f_B=1.14$ and $f_C=-44.53$ from Eq.~\ref{eq:eq05} (i.e. we do indeed have force balance in the bulk). 

We are now in a position to compare the force profiles that follow from the pressure tensor gradients with  those that follow from the chemical potential gradient. Figure~\ref{fig:fig03} (c) shows the profiles of the volume force at $\rho_C \sim 0.02$ with $\nabla \rho_C=1.0$. As can be seen from the figure,  the two expressions for the surface force (Eq.~\ref{eq:eq02} and Eq.~\ref{eq:eq05}) produce significantly different results at the interface. Not surprisingly, the forces calculated from  the chemical potentials (Eq.~\ref{eq:eq05}),  are concentrated where there is an excess of solute  [Fig.~\ref{fig:fig02}(b)]. However, the forces calculated by using the local pressure tensors computed via the Irving-Kirkwood and virial definitions (Eq.~\ref{eq:eq02}), extend over larger distances and vary in sign. The Irving-Kirkwood and virial definitions lead to  force profiles that are very similar. We verified that the integrated Marangoni force is effectively the same for all methods used:  $-4.7\pm0.1$ for the chemical potential, $-4.8\pm0.1$ for the virial and $-4.8\pm0.1$ for the Irving-Kirkwood, respectively. Moreover, these values agree  with  the surface tension gradient calculated from the values of surface tension at different concentrations, which is $-4.8\pm0.1$ at $\rho_C \sim 0.02$ with $\nabla \rho_C =1.0$ (calculated via $(\partial \gamma / \partial \rho_C ) \nabla \rho_C$). 
\begin{figure}[tb]
	\begin{center}
		 \includegraphics[width=0.45\textwidth]{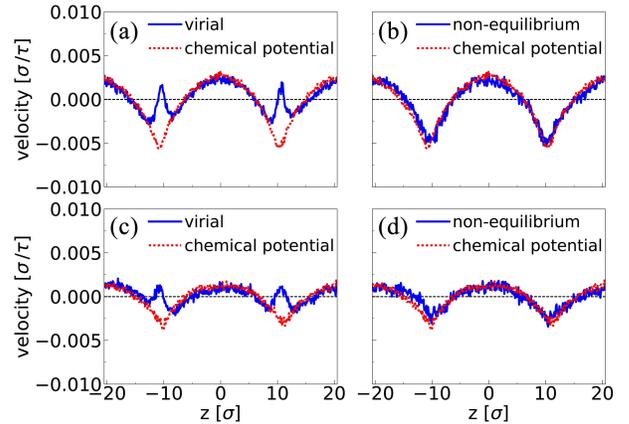}
		\caption{The velocity profiles along $z$ from different methods at $\nabla \rho_C=0.0010$ (a-b) and at $\nabla \rho_C=0.0005$ (c-d). The horizontal dashed line corresponds to the velocity of zero.}
		\label{fig:fig04}
	\end{center}
\end{figure}

The fact that pressure-tensor and chemical potential routes lead to different force profiles implies that they would result in different flow profiles. At most, one can be correct. To test this, we applied the force profiles that we computed to the fluid mixture at $\rho_C\sim0.02$ and measured the flow profile as a function of $z$ for fixed  $\nabla \rho_C$. The  Irving-Kirkwood and virial definitions lead to very similar results for the surface force, and hence we show only the virial flow profile. Figure~\ref{fig:fig04}(a) shows the predicted velocity profiles at $\nabla \rho_C=0.001$. We see that although the velocity profiles are very similar in the bulk, they are significantly different near the interface. For the sake of comparison,  the velocity profile obtained in a non-equilibrium MD simulation with an imposed concentration gradient of $\nabla \rho_C=0.001$ was determined in a region with $-10<x<10$ ($x=0$ at the center of the box). The result is shown in Fig.~\ref{fig:fig04}(b).  We see that the velocity profile that follows from the direct simulation differs markedly from the one obtained from the pressure tensor gradients. However, it agrees quite well with the predictions based on the chemical-potential gradient calculations. The same results were found  at $\nabla \rho_C=0.0005$ [Fig.~\ref{fig:fig04} (c) and (d)]. 
 
This finding is interesting because it indicates that the use of local pressure-tensor gradients leads to incorrect prediction of the Marangoni flow profile near the interface, even though the velocity in the bulk is still reliable. The latter finding is consistent with our previous work~\cite{ganti2017molecular}, which showed that bulk thermo-osmotic flow computed via local pressure gradients agrees well with the flow predicted by its reciprocal mechano-caloric coefficient.
Our results suggest that the chemical potential route, which is anyway the simplest, should be the preferred route to compute microscopic Marangoni flows.

In their original hydrodynamic formulation of the stress tensor~\cite{Irving1950}, Irving and Kirkwood note that a boundary or interface can cause the stress to depend on gradients of the pairwise atomic density, which can be neglected in the standard Irving-Kirkwood expression for fluids in the absence of gradients. The present work provides evidence that the problem hinted at by Irving and Kirkwood indeed becomes important in a gradient near an interface, as the potential part of the stress tensor then depends not only on the distance of two points between which a force acts, but also on the absolute coordinates of these points.



\begin{acknowledgments}
This work was supported by the European Union through ETN NANOTRANS grant 674979 MSC  the Sackler Fund. YL would like to acknowledge the hospitality of the Chemistry Department of the University of Cambridge.
\end{acknowledgments}

\bibliography{text}

\end{document}